\documentclass[a4paper,twocolumn,prb,showpacs,preprintnumbers]{revtex4-1}
\usepackage[T1]{fontenc}
\usepackage[latin9]{inputenc}
\usepackage{array}
\usepackage{multirow}
\usepackage{graphicx}
\usepackage{amstext}
\makeatletter

\pdfpageheight\paperheight
\pdfpagewidth\paperwidth

\providecommand{\tabularnewline}{\\}

\@ifundefined{textcolor}{}
{%
 \definecolor{BLACK}{gray}{0}
 \definecolor{WHITE}{gray}{1}
 \definecolor{RED}{rgb}{1,0,0}
 \definecolor{GREEN}{rgb}{0,1,0}
 \definecolor{BLUE}{rgb}{0,0,1}
 \definecolor{CYAN}{cmyk}{1,0,0,0}
 \definecolor{MAGENTA}{cmyk}{0,1,0,0}
 \definecolor{YELLOW}{cmyk}{0,0,1,0}
}

\usepackage[titletoc]{appendix}

\def\
\usepackage{babel}

\makeatother

\begin{document}

\title{Model Hamiltonian for topological Kondo insulator SmB$_{6}$}

\author{Rui Yu$^{1}$, HongMing Weng$^{2,3}$, Xiao Hu$^{1}$, Zhong Fang$^{2,3}$ and  Xi Dai$^{2,3}$}

\address{$^{1}$ International Center for Materials Nanoarchitectonics (WPI-MANA),
National Institute for Materials Science, Tsukuba 305-0044, Japan}

\address{$^{2}$ Beijing National Laboratory for Condensed Matter Physics,
and Institute of Physics,Chinese Academy of Sciences, Beijing 100190,
China}
\address{$^{3}$ Collaborative Innovation Center of Quantum Matter, Beijing
100190, China}

\email{Hu.Xiao@nims.go.jp}
\email{daix@aphy.iphy.ac.cn}
\date{\today}

\pacs{73.43.-f, 73.22.-f, 71.70.Ej, 85.75.-d}
\begin{abstract}
Starting from the k$\cdot$p method
in combination with first-principles calculations, we systematically
derive the effective Hamiltonians that capture the low energy band
structures of recently discovered topological Kondo insulator SmB$_{6}$.
Using these effective Hamiltonians we can obtain both the energy dispersion 
and the spin texture of the topological surface states, which can be detected by further
experiments. 
\end{abstract}
\maketitle

\section{Introduction}
Searching for new topological insulators (TI) has become an active
research field in condensed matter physics\cite{Qi_RMP:2011,Hasan_RMP:2010}.
A topological insulator has insulating and topologically non-trivial
bulk band structure giving rise to robust Dirac like surface states,
which are protected by time reversal symmetry and have the spin-momentum
locking feature. Such topological surface states have several remarkable
properties. For example, the suppression of back scattering and localization
on the TI surface\cite{roushan_nobackscattering_2009,alpichshev_stm_2010,
zhang_experimental_2009,xia_observation_2009,beidenkopf_spatial_2011}.
Furthermore, if superconductivity is induced on the surface of TIs
via proximity effects, the Majorana bound states can be induced \cite{fu_TI_Majorana_2008,santos_TI_Majorana_2010,qi_TI_Majorana_2010}.
These novel properties make topological insulator a promising platform
for the design of spintronics devices and future quantum computing
applications\cite{moore_birth_2010}.

Recently the mixed valence compound SmB$_{6}$ has been proposed
to be topological insulator and attracts lots of research interests\cite{Dzero_PRL_2009,
Dzero_prb_2012,Dzero_prl_2013,sunkai_2013,Dzero_JETP_2013,lu_smb6_2013}.
Unlike the other well studied topological insulator materials,
i.e. the Bi$_{2}$Se$_{3}$ family, the strong correlation effects
in mixed valence TIs are crucial to understand the electronic structure
due to the partially filled 4f bands\cite{lu_smb6_2013,Weng_ybb6_2013,Deng_TKI_2013}. There are two main effects induced
by the on-site Coulomb interaction among the f-electrons, one is the
strong modification of the 4f band width, the other is the correction
to the effective spin-orbital coupling and crystal field\cite{Coleman_2013PRL,MSigrist_2014_TKI,Werner_TKI_2013}. As a consequence,
the band inversion in the modified band structure happens between
the 5d and 4f band (with total angular momentum j=5/2) around the
three X points in the Brillouin Zone. Unlike the situation in Bi$_{2}$Se$_{3}$
family, where the band inversion happens between two bands both with
the p character and similar band width, the band inversion in SmB$_{6}$
happens between two bands with the band widths differing by orders,
which leads to very unique low energy electronic structure. Since
the band inversion happens at the Zone boundary (X points), which
project to three different points in surface Brillouin Zone leading
to three different Dirac points on generic surfaces.

Experimentally, the first evidence of topological surface states has been
found  by transport measurements, and then by angle resolved
photo emission spectroscopy (ARPES), Scanning Tunneling Spectroscopy (STS), quantum oscillation
magneto-resistance measurements{\cite{smb6_arpes_Frantzeskakis_2013,
smb6_ss_conduct_2013,smb6_arpes_xunan_2013,
smb6_qoscillation_2013,smb6_arpes_Neupane_2013,
smb6_Transport_2013,smb6_arpes_fengdl_2013,
smb6_WAL_2013,smb6_Yee_2013,exp_TKI_2013,exp_ss_PRX_2013,
SmB6_stm_2014PRL,ARPES_smb6_review_2013}.
Unlike the electronic structures in large energy scale, which is
mainly determined by the local atomic physics, the topological nature
of the electronic structure can be fully described by the quasi-particle
structure only, whose form can be determined from the symmetry
principles. In the present paper, we will construct a k$\cdot$p  model
capturing the full topological and symmetry features of the low energy
quasi-particle  structure, which leads to topological surface
states with the renormalized Fermi velocities. All the symmetry allowed
terms in the above k$\cdot$p model have been obtained by
fitting with the band structure obtained by the LDA+Gutzwiller calculation
introduced in a previous paper\cite{lu_smb6_2013}. Such a analytical model gives a
clear theoretical description for the quasi-particle
structure of SmB$_{6}$, which can be widely used in the further studies.

The organization of the present paper is as follows. In Sec.\ref{sec:Crystal-structure-and_band_structure},
we present the crystal structure and band structure of SmB$_{6}$.
Then we construct the effective models to describe the bulk band structure
for this material from the symmetry considerations in Sec.\ref{sec:Model-Hamiltonian-for_SmB}.
Furthermore we calculate surface states and the spin texture on the
the (001) surface based on our model Hamiltonian and show that it
is consistent with the tight-binding calculation results. Conclusions
are given in the end of this paper.

\section{Crystal structure and band structure\label{sec:Crystal-structure-and_band_structure}}

In this section we first describe the crystal structure of the SmB$_{6}$
and then discuss the nontrivial topological bulk band structure of
it.

\textbf{\textit{crystal structure:}} SmB$_{6}$ has the CsCl-type
crystal structure with $Pm\bar{3}m$ space group. The Sm ions are
located at the corner and B$_{6}$ octahedron are located at the body
center of the cubic lattice as shown in Figure \ref{fig:CS_and_BZ}(a).
The corresponding bulk and projected Brillouin zone (BZ) of (001)
surface for SmB$_{6}$ are shown in Fig.\ref{fig:CS_and_BZ}(b).

\begin{figure}
\includegraphics[scale=0.4]{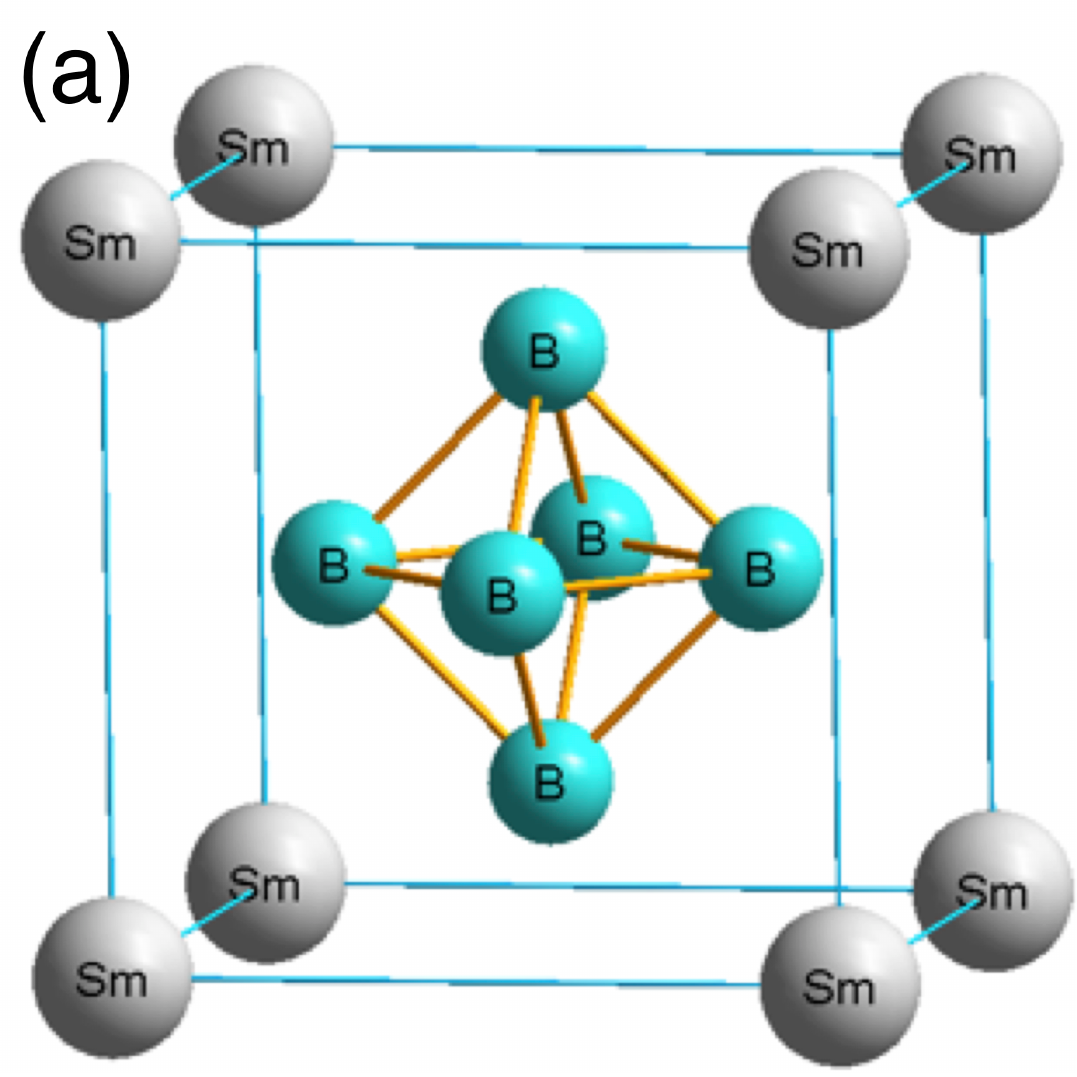} \includegraphics[scale=0.61]{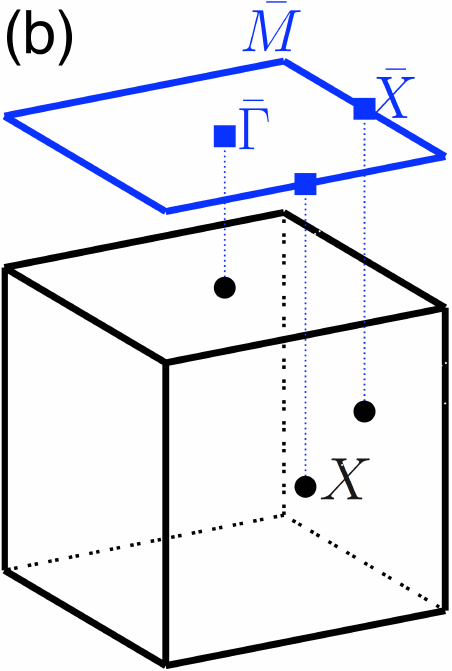}
\protect\caption{\label{fig:CS_and_BZ} (Color online) (a) The CsCl-type structure
of SmB$_{6}$with $Pm\bar{3}m$ space group. Sm ions and B$_{6}$
octahedron are located at the corner and center of the cubic lattice
respectively. (b) The bulk Brillouin zone of SmB$_{6}$ (black cubic)
and its projection onto the (001) (blue square). The $X$ points (black
dots) in the bulk BZ are projected to $\bar{\Gamma}$ and $\bar{X}$
points (blue square points ) in the (001) surface BZ.}
\end{figure}

\textbf{\textit{electronic structure:}} Previous electronic structure
studies find that in SmB$_{6}$ the band inversion happens between
Sm 4f bands with total angular momentum $j=5/2$ and one of the 5d
bands, which leads to fractional occupation in 4f $j=5/2$ orbitals
or \textquotedbl{}non-integer chemical valence\textquotedbl{} \cite{beaurepaire_MVS_1990,cohen_SmB6_Kondo_1970,eibschutz_SmB6_1972,chazalviel_MVs_1976}.
Since the band inversion happens at three X points in the BZ, where
the 4f and 5d states have opposite parities, the Z$_{2}$ topological
non-trivial band structure is  formed\cite{kane_z2_2005,Fu_kane_z2_2006}.
In order to include the strong Coulomb interaction among the f electrons,
we implement the local density approximation in density functional
theory with the Gutzwiller variational method (LDA+Gutzwiller) and
apply it to calculate the renormalized band structure of SmB$_{6}$.

Here we briefly introduce the LDA+Gutzwiller method, for detailed descriptions
please refer to Ref.{[}\onlinecite{DengXY_GW, lu_smb6_2013}{]}. The
LDA+Gutzwiller method combines the LDA with Gutzwiller variational
method, which takes care of the strong atomic feature of the f-orbitals
in the ground state wave function. In this method, we implement the
single particle Hamiltonian obtained by LDA with on site interaction
terms describing the atomic multiplet features, which can be written
as, 
\begin{equation}
H=H_{LDA}+H_{int}+H_{DC}\label{eq:H_tot}
\end{equation}
where $H_{LDA}$, $H_{int}$ and $H_{DC}$ represent the LDA Hamiltonian,
on-site interaction and the double counting terms respectively\cite{Kotliar:2006}.
The LDA Hamiltonian can be expressed in a tight binding form by constructing
the projected Wannier functions for both 5d and 4f bands\cite{wannier_2010,wannier_2007,wannier_2008}.The
on-site interactions can be described in terms of Slater integrals
as introduced in detail in the previous paper\cite{lu_smb6_2013}.
The double counting term $H_{DC}$ subtracts the correlation energy
already included in LDA calculation. Within the Gutzwiller approximation,
an effective Hamiltonian $H_{eff}$ describing the quasi-particle
band structure can be obtained, which describes the low energy dynamics
including the topological surface states\cite{lu_smb6_2013}. To further
study the low energy physics of SmB$_{6}$, i.e. the behavior of the
surface states, a simple k$\cdot$p model Hamiltonian will be very
useful. In the next section, we will construct such a model by expanding
the $H_{eff}$ near the three X points.


\section{Model Hamiltonian for SmB$_{6}$ \label{sec:Model-Hamiltonian-for_SmB}}

In this section, we will systematically derive the effective Hamiltonian
near $X$ points based on k$\cdot$p theory combined with the results
of first-principle calculations. We only give the effective Hamiltonian
at $X_{1}=(0,0,\frac{1}{2})$ point. The effective Hamiltonian at
the other two $X$ points can be obtained by acting $C_{4x}$ or $C_{4y}$
rotation operations on the Hamiltonian at $X_{1}$. Using the symmetry
group at the $X$ point we can construct the effective k$\cdot$p
model near this point and all the parameters used in such a model
can be obtained by fitting to the renormalized band structure obtained
by LDA+Gutzwiller.

The k$\cdot$p Hamiltonian is obtained from our one-partial effective
Hamiltonian 
\begin{equation}
H_{eff}\psi_{n,\mathbf{k}}(\mathbf{r})=E_{n,\mathbf{k}}\psi_{n,\mathbf{k}}(\mathbf{r}),\label{eq: Schrodinger EQ}
\end{equation}
where  $\psi_{n,\mathbf{k}}=e^{i\mathbf{k}\cdot\mathbf{r}}u_{n,\mathbf{k}}(\mathbf{r})$ are the Bloch
wave functions and the effective Hamiltonian only consists of the
kinetic-energy operator, a local periodic crystal potential, and the
spin-orbit interaction term: 
\begin{equation}
H_{eff}=\frac{p^{2}}{2m}+V(\mathbf{r})+\frac{\hbar}{4m_{0}^{2}c^{2}}(\mathbf{\sigma}\times\nabla V)\cdot\mathbf{p}.\label{eq: H_eff}
\end{equation}
In terms of the cellular functions $u_{n,\mathbf{k}}(\mathbf{r}),$ Eq.(\ref{eq: Schrodinger EQ}) becomes 
\begin{eqnarray}
H_{\mathbf{k}}u_{n,\mathbf{k}}(\mathbf{r}) & \equiv & \bigg[\frac{p^{2}}{2m}+V(\mathbf{r})+\frac{\hbar}{4m_{0}^{2}c^{2}}(\sigma\times\nabla V)\cdot(\mathbf{p}+\hbar\mathbf{k})\nonumber \\
 & + & \frac{\hbar}{m}\mathbf{k}\cdot\mathbf{p}\bigg]u_{n,\mathbf{k}}(\mathbf{r})=\epsilon_{n,\mathbf{k}}u_{n,\mathbf{k}}(\mathbf{r}),\label{eq:Hu=Eu}
\end{eqnarray}
where $\epsilon_{n,\mathbf{k}}\equiv E_{n,\mathbf{k}}-\frac{\hbar^{2}k^{2}}{2m}$.
Expanding the above Hamiltonian at given high symmetry point $k_{0}$, the eigen-equation at $k_{0}+k$
can be obtained by
\begin{equation}
\bigg[H_{\mathbf{k}_{0}}+\frac{\hbar}{m}\mathbf{k}\cdot\mathbf{p}\bigg]u_{n,\mathbf{k}_{0}+\mathbf{k}}(\mathbf{r})=\epsilon_{n,\mathbf{k}_{0}+\mathbf{k}}u_{n,\mathbf{k}_{0}+\mathbf{k}}(\mathbf{r}),\label{eq: H(k0+k)}
\end{equation}
where we have ignored the k-dependent spin-orbit term, which is usually
much small. Once $E_{n,\mathbf{k}_{0}}$ and $u_{n,\mathbf{k}_{0}}$are
known, the function $u_{n,\mathbf{k}_{0}+\mathbf{k}}(\mathbf{r})$
can be obtained by treating the term $H_{kp}=\frac{\hbar}{m}\mathbf{k}\cdot\mathbf{p}$
in Eq.(\ref{eq: H(k0+k)}) as a perturbation. It is more convenient
to rewrite the perturbation as 
\begin{equation}
H_{kp}=\frac{\hbar}{m}\mathbf{k}\cdot\mathbf{p}=\frac{\hbar}{2m}(k_{+}p_{-}+k_{-}p_{+})+\frac{\hbar}{m}k_{z}p_{z},\label{eq:H_kp}
\end{equation}
where the operator $p_{\pm}=p_{x}\pm ip_{y}$ and $k_{\pm}=k_{x}\pm ik_{y}$.

In the SmB$_{6}$ system, we  expand the $H_{eff}$ near the
three X points. We chose the k$\cdot$p basis function at X point as 
\begin{eqnarray}
 &  & 
 \left| \frac{3}{2},\frac{3}{2}\right\rangle_{d},
 \left| \frac{3}{2},-\frac{3}{2}\right\rangle_{d},
 \left|\frac{5}{2},\frac{5}{2}\right\rangle_{f},
 \left|\frac{5}{2},-\frac{5}{2}\right\rangle_{f}\nonumber \\
 &  & 
 \left|\frac{5}{2},\frac{3}{2}\right\rangle_{f},
 \left|\frac{5}{2},-\frac{3}{2}\right\rangle_{f},
 \left|\frac{5}{2},\frac{1}{2}\right\rangle_{f},
 \left|\frac{5}{2},-\frac{1}{2}\right\rangle_{f},
 \label{eq:kp basis}
\end{eqnarray}
which can well describe the orbital characters for the eigenstates near the Fermi energy. 
We can then project the k$\cdot$p
Hamiltonian into above basis and the matrix elements of $H_{kp}$
are constrained by the crystal symmetries at $X$ point. The little group
at $X$ point is $D_{4h}$,  which contains the following symmetry operations:

(1) fourfold rotation along the $z$ direction $\hat{C}_{4z}$ =$e^{-i\frac{2\pi}{4}\hat{J}_{z}}$,
where $\hat{J}_{\alpha}$ ($\alpha=x,y,z$) is the operator for the
$\alpha$ component of the total angular momentum.

(2) inversion symmetry $\hat{P}=I_{2}\oplus-I_{6}$, where $I_{m}$
is the m$\times$m identity matrix.

(3) time reversal symmetry $\hat{T}$=$\Theta K=e^{-i\pi\hat{J}_{y}}K$,
where $K$ is the complex conjugate operator.

(4) twofold rotation along $y$ direction $\hat{C}_{2y}=e^{-i\pi\hat{J}_{y}}$
.

The symmetry operation can help us to reduce the independent parameters that appear in the $k\cdot p$ 
Hamiltonian. For example, considering the
rotation $C_{4z}$  around the $z$ direction, we have 
\begin{equation}
_{d}\left\langle\frac{3}{2},\frac{3}{2}\right |p_{-}\left |\frac{5}{2},\frac{5}{2}\right \rangle_{f}
={}_{d}\left \langle\frac{3}{2},\frac{3}{2}\right |C_{4z}^{\dagger}C_{4z}p_{-}C_{4z}^{\dagger}C_{4z}\left |\frac{5}{2},\frac{5}{2}\right \rangle_{f}.\label{eq:c4_invariant}
\end{equation}
Since $_{d}\langle\frac{3}{2},\frac{3}{2}|C_{4z}^{\dagger}={}_{d}\langle\frac{3}{2},\frac{3}{2}|e^{i\frac{2\pi}{4}\frac{3}{2}}$,
$C_{4z}p_{\pm}C_{4z}^{\dagger}=e^{\mp i\frac{2\pi}{4}}p_{\pm}$, and
$C_{4z}|\frac{5}{2},\frac{5}{2}\rangle_{f}=e^{-i\frac{2\pi}{4}\frac{5}{2}}|\frac{5}{2},\frac{5}{2}\rangle_{f}$,
we get that $c_{1}\equiv\frac{\hbar}{2m}{}_{d}\langle\frac{3}{2},\frac{3}{2}|p_{-}|\frac{5}{2},\frac{5}{2}\rangle_{f}$
is invariant under $C_{4z}$ rotation and can be non-zero. While $_{d}\langle\frac{3}{2},\frac{3}{2}|p_{+}|\frac{5}{2},\frac{5}{2}\rangle_{f}$
get a minus sign under $C_{4z}$ rotation, which means it must vanish. Following
the same procedge, we can get $c_{1}^{\prime}\equiv\frac{\hbar}{2m}{}_{d}\langle\frac{3}{2},-\frac{3}{2}|p_{+}|\frac{5}{2},-\frac{5}{2}\rangle_{f}$
is finite. When considering the 2-fold rotation along the $y$ direction $C_{2y}$,
we get the relation between $c_{1}$ and $c_{1}^{\prime}$ as 
\begin{eqnarray}
c_{1} & = & \frac{\hbar}{2m}{}_{d}\langle\frac{3}{2},\frac{3}{2}|p_{-}|\frac{5}{2},\frac{5}{2}\rangle_{f}\nonumber \\
 & = & \frac{\hbar}{2m}{}_{d}\langle\frac{3}{2},\frac{3}{2}|C_{2y}^{\dagger}C_{2y}p_{-}C_{2y}^{\dagger}C_{2y}|\frac{5}{2},\frac{5}{2}\rangle_{f}\nonumber \\
 & = & -\frac{\hbar}{2m}{}_{d}\langle\frac{3}{2},-\frac{3}{2}|p_{+}|\frac{5}{2},-\frac{5}{2}\rangle_{f}=-c_{1}^{\prime}.\label{eq:c1=-c1p}
\end{eqnarray}
Due to the time-reversal symmetry, the $c_{1}$ can be chosen
to be real. Similar considerations can be used to calculate all matrix
elements and we obtain the effective
Hamiltonian, which is invariant under all symmetry operations at $X$ point up to the first order of $k$, 
\begin{equation}
H_{X}=\left(\begin{array}{cccccccc}
\epsilon_{d} & 0 & c_{1}k_{+} & c_{2}k_{z} & c_{3}k_{z} & c_{4}k_{+} & c_{5}k_{-} & 0\\
 & \epsilon_{d} & c_{2}k_{z} & -c_{1}k_{-} & -c_{4}k_{-} & c_{3}k_{z} & 0 & -c_{5}k_{+}\\
 &  & \epsilon_{f_{5}} & 0 & 0 & d_{1} & 0 & 0\\
 &  &  & \epsilon_{f_{5}} & d_{1} & 0 & 0 & 0\\
 &  &  &  & \epsilon_{f_{3}} & 0 & 0 & 0\\
 &  & \dagger &  &  & \epsilon_{f_{3}} & 0 & 0\\
 &  &  &  &  &  & \epsilon_{f_{1}} & 0\\
 &  &  &  &  &  &  & \epsilon_{f_{1}}
\end{array}\right),\label{eq:H_SmB6}
\end{equation}
where $\epsilon_{d}=D+D_{xy}(k_{x}^{2}+k_{y}^{2})+D_{z}k_{z}^{2}$, 
$\epsilon_{f_{i}}=F_{i}+F_{i,xy}(k_{x}^{2}+k_{y}^{2})+F_{i,z}k_{z}^{2}$
($i=5,3,1$) and $k_{\pm}=k_{x}\pm ik_{y}$. The parameters are listed
in Table.\ref{tab:kp_paras}. The fitted energy dispersion for SmB$_{6}$
is plotted in Fig.\ref{fig:SmB6_3D}. It shows that our model Hamiltonian
with eight bands captures the main features of the band dispersion
near $X$ point.

\begin{table}[h]
\protect\caption{\label{tab:kp_paras} Parameters in our model Hamiltonian
fitted for SmB$_{6}$, where the unit of energy is eV and the unit
of length is the lattice constant.}
\begin{tabular}{cccccc}
\hline 
$D$  & $D_{xy}$  & $D_{z}$  & $F_{1}$  & $F_{1,xy}$  & $F_{1,z}$\tabularnewline
-1.5698  & 29.9233  & 18.2502  & -0.0532  & -0.0020  & -0.0300\tabularnewline
\hline 
$F_{3}$  & $F_{3,xy}$  & $F_{3,z}$  & $F_{5}$  & $F_{5,xy}$  & $F_{5,z}$\tabularnewline
0.0163  & -0.5776  & -0.4476  & -0.0164  & -0.1543  & -0.1543\tabularnewline
\hline 
$c_{1}$  & $c_{2}$  & $c_{3}$  & $c_{4}$  & $c_{5}$  & $d_{1}$\tabularnewline
-0.2787  & -0.318  & -0.502  & -0.5960  & 0.0021  & -0.0132\tabularnewline
\hline 
\end{tabular}
\end{table}
\begin{figure}
\begin{centering}
\includegraphics[scale=0.4]{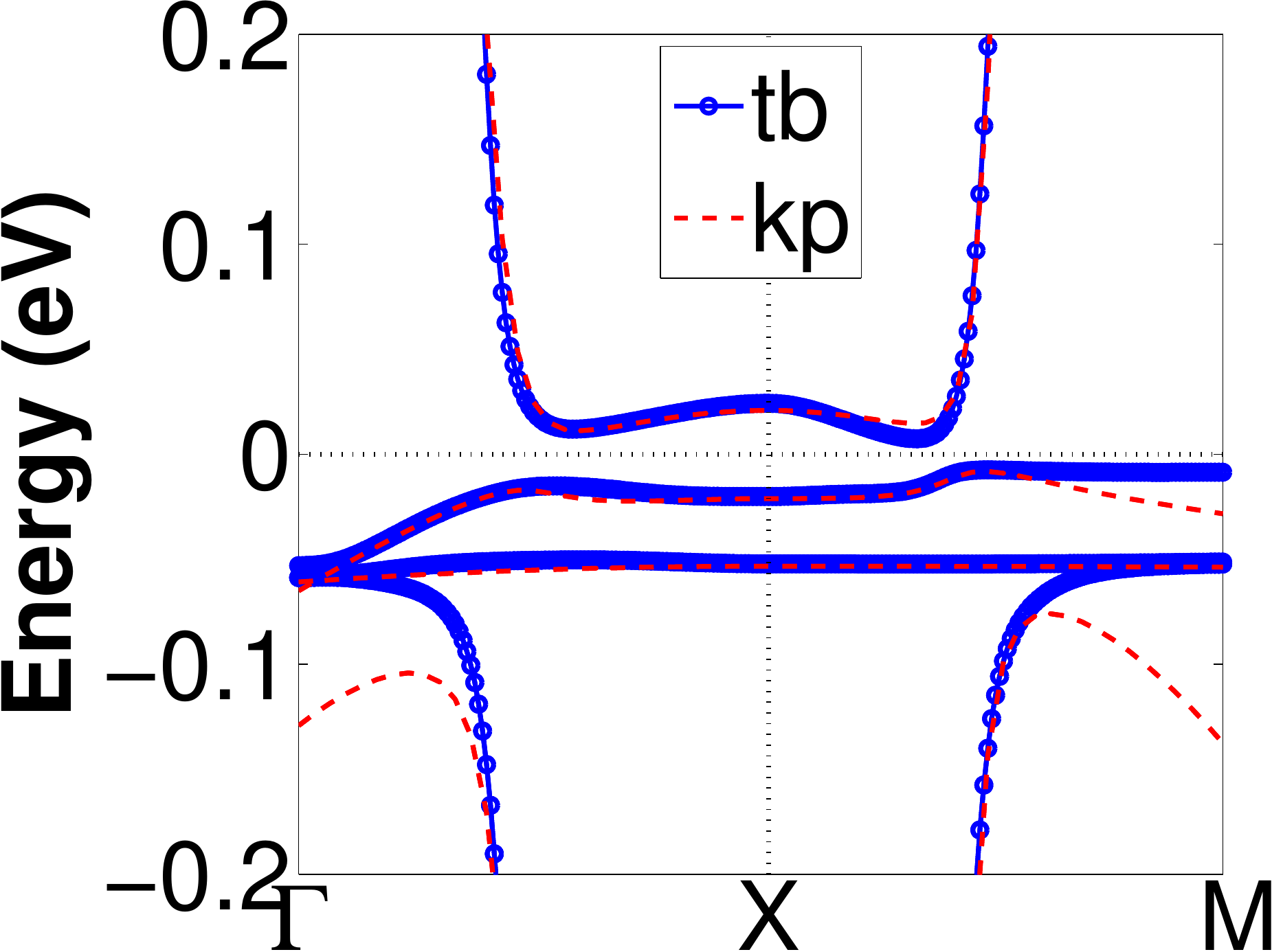} 
\par\end{centering}

\protect\protect\caption{\label{fig:SmB6_3D}The energy dispersion obtained from our model
Hamiltonian with 8 bands (thin red line) is compared with that from
tight-binding calculations (thick blue line). Here $M$ to $X$ lines
represents the dispersion along the $k_{x}$ direction while the $X$
to $\Gamma$ lines is for the $k_{z}$ direction..}
\end{figure}

An important physical consequence of the non-trivial topological band
structure is the existence of Dirac like surface states with chiral
spin texture. The $X$ points in the bulk BZ are projected to $\bar{\Gamma}$
and $\bar{X}$ points in the (001) surface BZ as shown in Fig.\ref{fig:CS_and_BZ}(b).
To study the surface state and the spin texture near $\bar{\Gamma}$
point, we consider a thick slab limited in $z\in[-d/2,d/2]$ with
open boundary conditions, where $d$ is the thickness of the slab
in $z$ direction. Now $k_{z}$ is not a good quantum number which
should be replaced by $-i\partial_{z}$. The eigenwave function will
be given by $\psi(k_{x},k_{y},z)$, which can be expanded using basis
\{$\varphi_{n}(z)=\sqrt{2/d}\; sin[n\pi(z+\frac{d}{2})/d]$\} $(n=0,1,2,3,...)$.
The Hamiltonian for the slab structure is written as $H_{\bar{\Gamma},mn}^{slab}(k_{x},k_{y})=\langle\varphi_{m}(z)|H_{s}(k_{x},k_{y},-i\partial_{z})|\varphi_{n}(z)\rangle$.
The surface states near $\bar{\Gamma}$ point can be calculated directly
from $H_{\bar{\Gamma}}^{slab}(k_{x},k_{y})$. For the surface states
and spin texture near $\bar{X}$ point, we can use the same method
but change the $z$ direction to $x$ direction. The calculated surface
states near $\bar{\Gamma}$ and $\bar{X}$ points are shown in Fig.\ref{fig:SmB6_2D}(b)
and compared with the results from tight-binding calculation Fig.\ref{fig:SmB6_2D}(a).
There are three Dirac cone like surface states. One located at $\bar{\Gamma}$
points, the other two located at two $\bar{X}$ points, which is different
with most of the known 3D topological insulators, such as B$i_{2}$Se$_{3}$.

\vspace{2mm}

\begin{figure}
\begin{centering}
\includegraphics[scale=0.36]{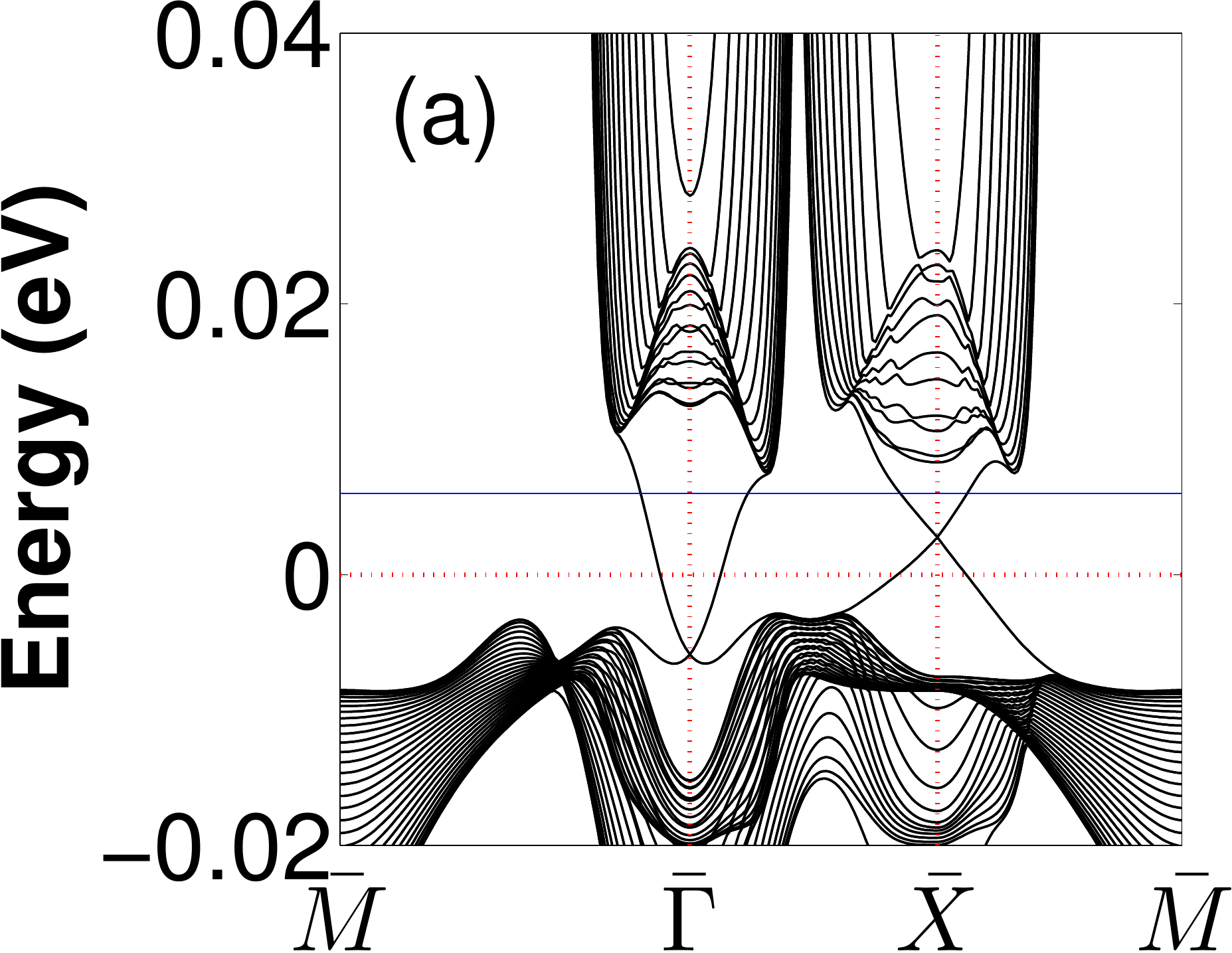} 
\par\end{centering}

\begin{centering}
\includegraphics[scale=0.36]{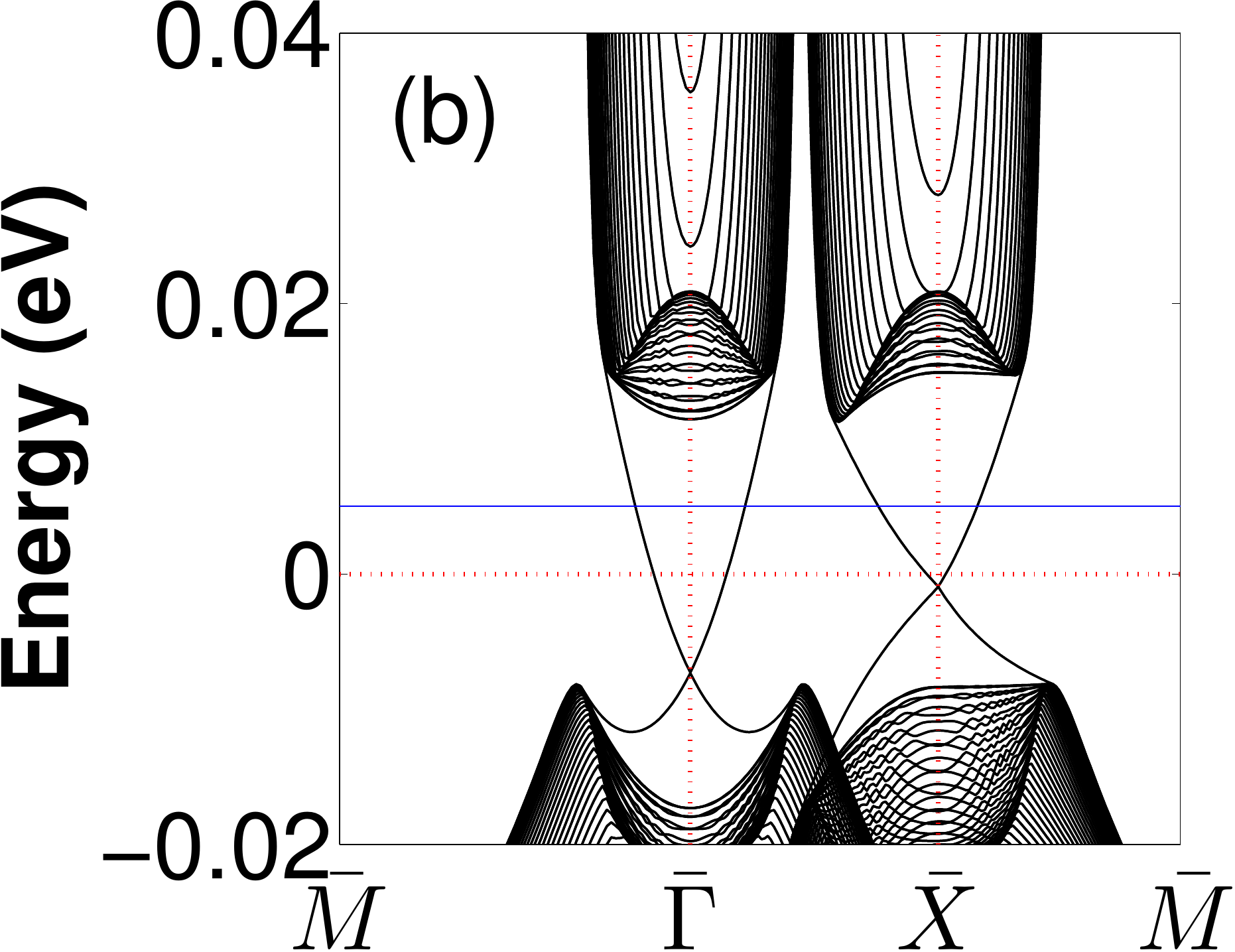} 
\par\end{centering}

\protect\protect\caption{\label{fig:SmB6_2D} The (001) direction surface states near $\bar{\Gamma}$
and $\bar{X}$ point, (a) from the tight-binding model; (b) from the
k$\cdot$p model results. The blue solid line indicates the energy
for the spin-texture calculations.}
\end{figure}

From $H_{\bar{\Gamma},\bar{X}}^{slab}$ we can further get the spin
texture of the surface states near $\bar{\Gamma}$ and $\bar{X}$
points. The spin texture at the energy of 6meV is shown in Fig.\ref{fig:SmB6_spin_texture},
which shows a strong spin-moment locking on the surface states. We
will discuss the spin texture from the symmetry consideration below.

\begin{figure*}
\begin{centering}
\includegraphics[scale=0.4]{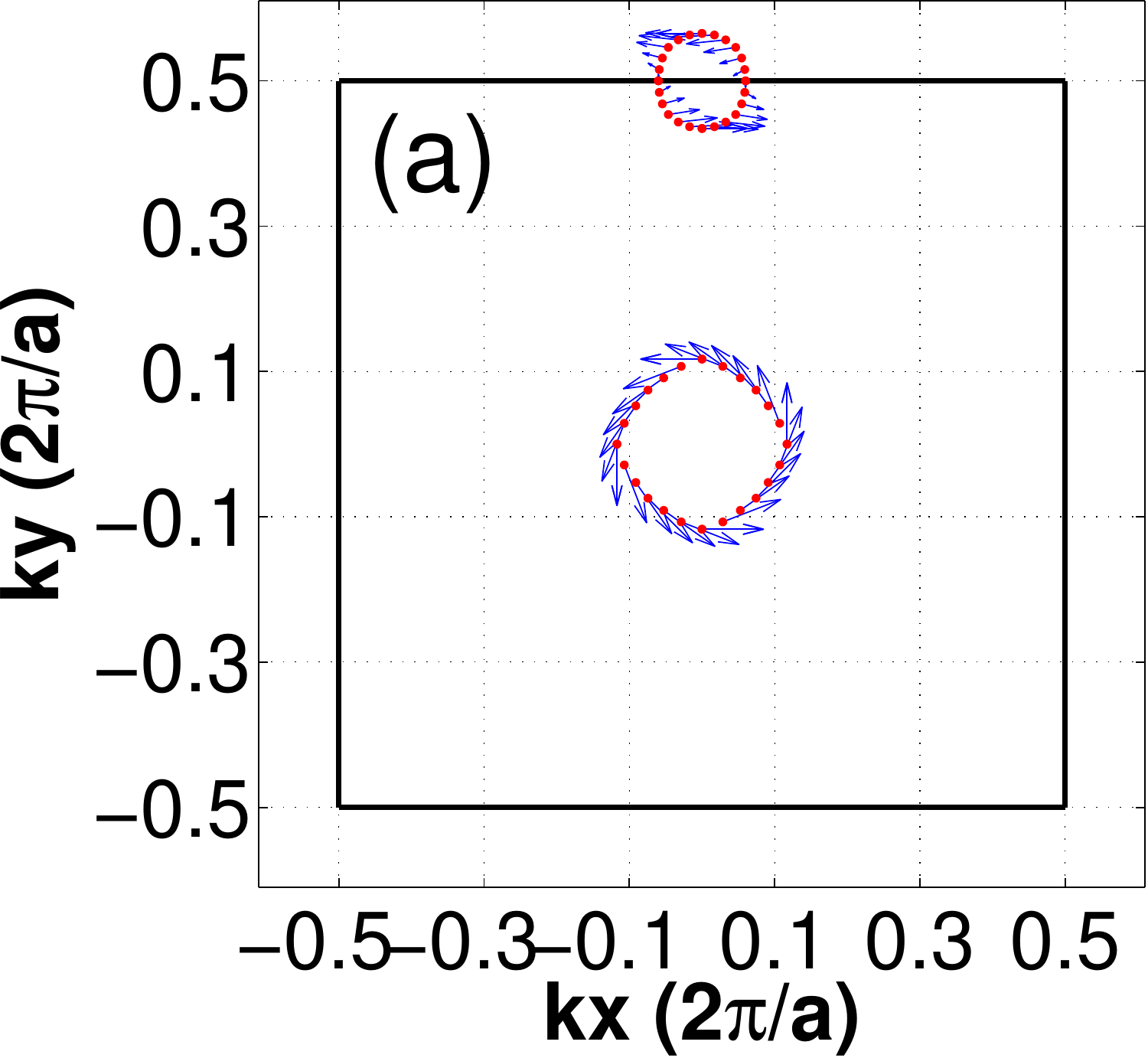}$\;\;$\includegraphics[scale=0.4]{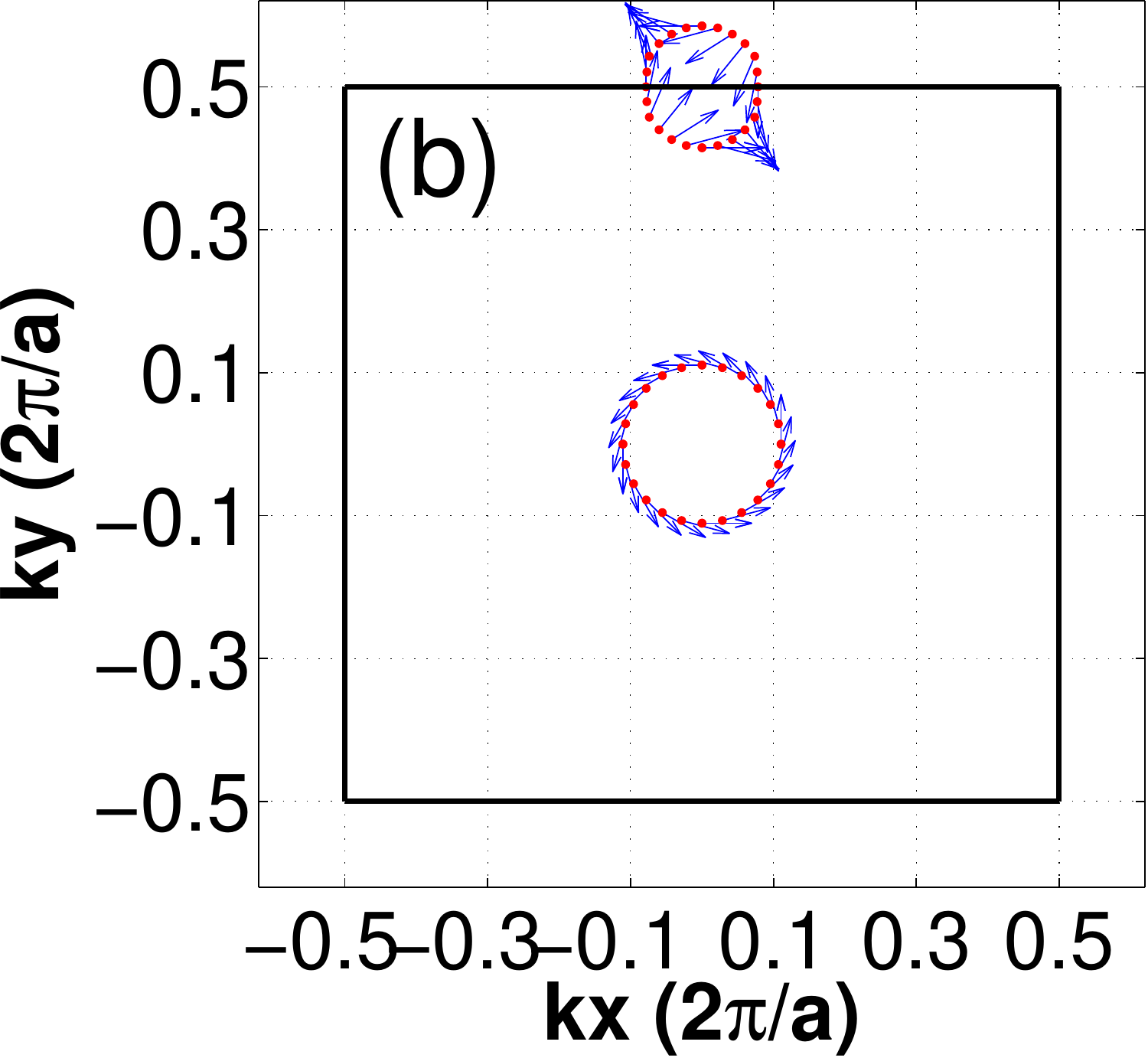} 
\par\end{centering}

\protect\protect\caption{\label{fig:SmB6_spin_texture} Spin texture on the (001) direction
surface states near $\bar{\Gamma}$ and $\bar{X}$ point, (a) from the
tight-binding model; (b) from the k$\cdot$p model results.}
\end{figure*}

To understand the spin orientation on the surface states, we can construct
the surface effective Hamiltonian at $\bar{\Gamma}$ and $\bar{X}$
points on the 2D projected surface BZ. On the 2D BZ, the little group
is $C_{4v}$ at $\bar{\Gamma}$ point and $C_{2v}$ at $\bar{X}$
point. The surface states at these two points are transformed as the
states with $j_{z}=\pm3/2$ under the symmetry operation in the little
group. Form the symmetry consideration, we can write down the effective
Hamiltonian for the surface states at $\bar{\Gamma}$ and $\bar{X}$
points respectively:

1) At $\bar{\Gamma}$ point: The Hamiltonian  satisfy the $C_{4v}$
symmetry and time-reversal symmetry must having the following form up
to the third order of $k$:
\begin{equation}
H_{\bar{\Gamma}}^{ss}(k)=c_{0}k^{2}+\bigg[i(c_{1}k_{+}+c_{2}k_{+}^{2}k_{-}+c_{3}k_{-}^{3})\hat{\sigma}_{+}+h.c.\bigg],\label{eq:H_ss_G}
\end{equation}
in the basis of \{$|j_{z}=\frac{3}{2}\rangle,|j_{z}=-\frac{3}{2}\rangle$\}.
Where $\hat{\sigma}_{\pm}=\hat{\sigma}_{x}\pm i\hat{\sigma}_{y}$,
$k_{\pm}=k_{x}\pm ik_{y}$ and $k^{2}=k_{x}^{2}+k_{y}^{2}$. The parameters
are listed in table.\ref{tab:Parameters-for-ss}

Here we should determine the spin operators $\hat{s}_{x,y,z}$ for
the surface model Hamiltonian at $\bar{\Gamma}$ point . We start
from the tight-binding Hamiltonian $H_{eff}$, which is written in
the tight-binding basis space expanded by $d$ and $f$ orbitals located
on Sm atoms. Based on this tight-binding model we can construct a
thick slab terminated in the (001) direction and calculate the surface
states $|\Psi_{\alpha}\rangle$, $\alpha=1,2$. The spin operator
$\mathbf{\hat{S}}$ for this slab system can be easily written in
the tight-binding basis. Then, we project the spin operator $\mathbf{\hat{S}}$
onto the surface states subspace. Finally, we find that the spin operators
for surface states are 
\begin{eqnarray}
s_{x}^{\alpha\beta} & = & \langle\Psi_{\alpha}|\hat{S}_{x}|\Psi_{\beta}\rangle=0.0953\sigma_{x}^{\alpha\beta} \nonumber \\
s_{y}^{\alpha\beta} & = &\langle\Psi_{\alpha}|\hat{S}_{y}|\Psi_{\beta}\rangle=-0.0953\sigma_{y}^{\alpha\beta}\nonumber\\
 s_{z}^{\alpha\beta} & = &\langle\Psi_{\alpha}|\hat{S}_{z}|\Psi_{\beta}\rangle=0.0638\sigma_{z}^{\alpha\beta}
\end{eqnarray}
The total angular momentum operators in the surface states subspace
can also be calculated as $\hat{j}_{x}=-0.6648\hat{\sigma}_{x}$,
$\hat{j}_{y}=0.6648\hat{\sigma}_{y}$ $\hat{j}_{z}=-0.5405\hat{\sigma}_{z}$. 

In order to predict or understand  properties of the surface
states under the external magnetic field, we give the Zeeman coupling terms for surface
states, which takes the following form
\begin{equation}
H_{\bar{\Gamma}}^{ss,Z}=\mu_{B}\sum_{\alpha,\beta}g_{\alpha\beta}{\hat\sigma}_{\alpha}B_{\beta},
\label{eq:H_zeeman}
\end{equation}
which is obtained by projecting the $H_{Zeeman}=\frac{\mu_{B}}{\hbar}(\hat{\mathbf{L}}+2\hat{\mathbf{S}})\cdot\mathbf{B}$
term into the surface states subspace. Here $\hat{\mathbf{L}}$ and
$\hat{\mathbf{S}}$ are the orbital angular momentum and spin operator
of the slab system. The the non-zero matrix elements of the g factor matrix for surface states at $\bar{\Gamma}$
point are listed in table.\ref{tab:Parameters-for-ss} 

2) At $\bar{X}$ point: As shown in Fig.(\ref{fig:SmB6_2D}), the
Dirac cone has a good linear dispersion. The surface effective Hamiltonian
which satisfying the $C_{2v}$ symmetry and time-reversal symmetry must
have the following form up to the second order of $k$ in the basis
of \{$|j_{z}=\frac{3}{2}\rangle,|j_{z}=-\frac{3}{2}\rangle$\}, 
\begin{equation}
H_{\bar{X}}^{ss}(\mathbf{k})=(\epsilon_{0}+a_{0}k^{2})\sigma_{0}+\big[i(a_{1}k_{+}+a_{2}k_{-})\hat{\sigma}_{+}+h.c.\big].\label{eq:H_ss_X}
\end{equation}
The spin operators for surface states are
\begin{equation}
\hat{s}_{x}=0.0687\hat{\sigma}_{x};\; 
\hat{s}_{y}=-0.1223\hat{\sigma}_{y};\;
\hat{s}_{z}=-0.1484\hat{\sigma}_{z}
\end{equation}
The total angular momentum operators in the surface states subspace
can also be calculated as $\hat{j}_{x}=-0.4201\hat{\sigma}_{x}$,
$\hat{j}_{y}=0.7870\hat{\sigma}_{y}$ $\hat{j}_{z}=0.9309\hat{\sigma}_{z}$,
The Zeeman term for surface states at $\bar{X}$ point is the same
as Eq.(\ref{eq:H_zeeman}) and the g factors are listed in table.\ref{tab:Parameters-for-ss}.

\begin{table}
\protect\protect\caption{Parameters for the surface states model. The unit of energy is eV
and the unit of length is the lattice constant.\label{tab:Parameters-for-ss}}

\begin{centering}
\begin{tabular}{ccccc}
\hline 
\multirow{4}{*}{$H_{\bar{\Gamma}}^{ss}$} & $c_{0}$  & $c_{1}$  & $c_{2}$  & $c_{3}$\tabularnewline
 & 0.64105  & 0.01524  & 1.1081  & -0.0516\tabularnewline
\cline{2-5} 
 & $g_{xx}$ & $g_{yy}$ & $g_{zz}$ & \tabularnewline
 & -0.5695 & 0.5695 & -0.4768 & \tabularnewline
\hline 
\multirow{4}{*}{$H_{\bar{X}}^{ss}$ } & $a_{0}$  & $a_{1}$  & $a_{2}$  & $\epsilon_{0}$\tabularnewline
 & 0.011276  & 0.003059  & -0.02322  & 0.008654\tabularnewline
\cline{2-5} 
 & $g_{xx}$ & $g_{yy}$ & $g_{zz}$ & \tabularnewline
 & -0.3514 & 0.6647 & 0.7825 & \tabularnewline
\hline 
\end{tabular}
\par\end{centering}
\end{table}


\section*{CONCLUSIONS}

To summarize, we have derived the model Hamiltonians around the three
X points for the 3D TI SmB$_{6}$ based on the first principles results
and the symmetry considerations. The bulk band structure, the surface states
on (001) surface and the spin texture of the surface states can be well
described by our model Hamiltonians. These effective Hamiltonians
could facilitate further investigations of similar intriguing materials.

\textit{Acknowledgments} 
This work was supported by the WPI Initiative on Materials Nanoarchitectonics, 
and Grant-in-Aid for Scientific Research under the Innovative Area "Topological 
Quantum Phenomena" (No.25103723), MEXT of Japan. H.M. Weng, Zhong Fang and X. Dai acknowledge the NSF of China and the 973 Program of China (No. 2011CBA00108 and No. 2013CB921700) for support.

 \bibliographystyle{apsrev}
\bibliography{ref}

\begin{thebibliography}{46}
\expandafter\ifx\csname natexlab\endcsname\relax\def\natexlab#1{#1}\fi
\expandafter\ifx\csname bibnamefont\endcsname\relax
  \def\bibnamefont#1{#1}\fi
\expandafter\ifx\csname bibfnamefont\endcsname\relax
  \def\bibfnamefont#1{#1}\fi
\expandafter\ifx\csname citenamefont\endcsname\relax
  \def\citenamefont#1{#1}\fi
\expandafter\ifx\csname url\endcsname\relax
  \def\url#1{\texttt{#1}}\fi
\expandafter\ifx\csname urlprefix\endcsname\relax\def\urlprefix{URL }\fi
\providecommand{\bibinfo}[2]{#2}
\providecommand{\eprint}[2][]{\url{#2}}

\bibitem[{\citenamefont{Qi and Zhang}(2011)}]{Qi_RMP:2011}
\bibinfo{author}{\bibfnamefont{X.-L.} \bibnamefont{Qi}} \bibnamefont{and}
  \bibinfo{author}{\bibfnamefont{S.-C.} \bibnamefont{Zhang}},
  \bibinfo{journal}{Reviews of Modern Physics} \textbf{\bibinfo{volume}{83}},
  \bibinfo{pages}{1057} (\bibinfo{year}{2011}).

\bibitem[{\citenamefont{Hasan and Kane}(2010)}]{Hasan_RMP:2010}
\bibinfo{author}{\bibfnamefont{M.~Z.} \bibnamefont{Hasan}} \bibnamefont{and}
  \bibinfo{author}{\bibfnamefont{C.~L.} \bibnamefont{Kane}},
  \bibinfo{journal}{Reviews of Modern Physics} \textbf{\bibinfo{volume}{82}},
  \bibinfo{pages}{3045} (\bibinfo{year}{2010}).

\bibitem[{\citenamefont{Roushan et~al.}(2009)\citenamefont{Roushan, Seo,
  Parker, Hor, Hsieh, Qian, Richardella, Hasan, Cava, and
  Yazdani}}]{roushan_nobackscattering_2009}
\bibinfo{author}{\bibfnamefont{P.}~\bibnamefont{Roushan}},
  \bibinfo{author}{\bibfnamefont{J.}~\bibnamefont{Seo}},
  \bibinfo{author}{\bibfnamefont{C.~V.} \bibnamefont{Parker}},
  \bibinfo{author}{\bibfnamefont{Y.~S.} \bibnamefont{Hor}},
  \bibinfo{author}{\bibfnamefont{D.}~\bibnamefont{Hsieh}},
  \bibinfo{author}{\bibfnamefont{D.}~\bibnamefont{Qian}},
  \bibinfo{author}{\bibfnamefont{A.}~\bibnamefont{Richardella}},
  \bibinfo{author}{\bibfnamefont{M.~Z.} \bibnamefont{Hasan}},
  \bibinfo{author}{\bibfnamefont{R.~J.} \bibnamefont{Cava}}, \bibnamefont{and}
  \bibinfo{author}{\bibfnamefont{A.}~\bibnamefont{Yazdani}},
  \bibinfo{journal}{Nature} \textbf{\bibinfo{volume}{460}},
  \bibinfo{pages}{1106} (\bibinfo{year}{2009}).

\bibitem[{\citenamefont{Alpichshev et~al.}(2010)\citenamefont{Alpichshev,
  Analytis, Chu, Fisher, Chen, Shen, Fang, and
  Kapitulnik}}]{alpichshev_stm_2010}
\bibinfo{author}{\bibfnamefont{Z.}~\bibnamefont{Alpichshev}},
  \bibinfo{author}{\bibfnamefont{J.~G.} \bibnamefont{Analytis}},
  \bibinfo{author}{\bibfnamefont{J.-H.} \bibnamefont{Chu}},
  \bibinfo{author}{\bibfnamefont{I.~R.} \bibnamefont{Fisher}},
  \bibinfo{author}{\bibfnamefont{Y.~L.} \bibnamefont{Chen}},
  \bibinfo{author}{\bibfnamefont{Z.~X.} \bibnamefont{Shen}},
  \bibinfo{author}{\bibfnamefont{A.}~\bibnamefont{Fang}}, \bibnamefont{and}
  \bibinfo{author}{\bibfnamefont{A.}~\bibnamefont{Kapitulnik}},
  \bibinfo{journal}{Physical Review Letters} \textbf{\bibinfo{volume}{104}},
  \bibinfo{pages}{016401} (\bibinfo{year}{2010}).

\bibitem[{\citenamefont{Zhang et~al.}(2009)\citenamefont{Zhang, Cheng, Chen,
  Jia, Ma, He, Wang, Zhang, Dai, Fang et~al.}}]{zhang_experimental_2009}
\bibinfo{author}{\bibfnamefont{T.}~\bibnamefont{Zhang}},
  \bibinfo{author}{\bibfnamefont{P.}~\bibnamefont{Cheng}},
  \bibinfo{author}{\bibfnamefont{X.}~\bibnamefont{Chen}},
  \bibinfo{author}{\bibfnamefont{J.-F.} \bibnamefont{Jia}},
  \bibinfo{author}{\bibfnamefont{X.}~\bibnamefont{Ma}},
  \bibinfo{author}{\bibfnamefont{K.}~\bibnamefont{He}},
  \bibinfo{author}{\bibfnamefont{L.}~\bibnamefont{Wang}},
  \bibinfo{author}{\bibfnamefont{H.}~\bibnamefont{Zhang}},
  \bibinfo{author}{\bibfnamefont{X.}~\bibnamefont{Dai}},
  \bibinfo{author}{\bibfnamefont{Z.}~\bibnamefont{Fang}}, \bibnamefont{et~al.},
  \bibinfo{journal}{Physical Review Letters} \textbf{\bibinfo{volume}{103}},
  \bibinfo{pages}{266803} (\bibinfo{year}{2009}).

\bibitem[{\citenamefont{Xia et~al.}(2009)\citenamefont{Xia, Qian, Hsieh, Wray,
  Pal, Lin, Bansil, Grauer, Hor, Cava et~al.}}]{xia_observation_2009}
\bibinfo{author}{\bibfnamefont{Y.}~\bibnamefont{Xia}},
  \bibinfo{author}{\bibfnamefont{D.}~\bibnamefont{Qian}},
  \bibinfo{author}{\bibfnamefont{D.}~\bibnamefont{Hsieh}},
  \bibinfo{author}{\bibfnamefont{L.}~\bibnamefont{Wray}},
  \bibinfo{author}{\bibfnamefont{A.}~\bibnamefont{Pal}},
  \bibinfo{author}{\bibfnamefont{H.}~\bibnamefont{Lin}},
  \bibinfo{author}{\bibfnamefont{A.}~\bibnamefont{Bansil}},
  \bibinfo{author}{\bibfnamefont{D.}~\bibnamefont{Grauer}},
  \bibinfo{author}{\bibfnamefont{Y.~S.} \bibnamefont{Hor}},
  \bibinfo{author}{\bibfnamefont{R.~J.} \bibnamefont{Cava}},
  \bibnamefont{et~al.}, \bibinfo{journal}{Nature Physics}
  \textbf{\bibinfo{volume}{5}}, \bibinfo{pages}{398} (\bibinfo{year}{2009}).

\bibitem[{\citenamefont{Beidenkopf et~al.}(2011)\citenamefont{Beidenkopf,
  Roushan, Seo, Gorman, Drozdov, Hor, Cava, and
  Yazdani}}]{beidenkopf_spatial_2011}
\bibinfo{author}{\bibfnamefont{H.}~\bibnamefont{Beidenkopf}},
  \bibinfo{author}{\bibfnamefont{P.}~\bibnamefont{Roushan}},
  \bibinfo{author}{\bibfnamefont{J.}~\bibnamefont{Seo}},
  \bibinfo{author}{\bibfnamefont{L.}~\bibnamefont{Gorman}},
  \bibinfo{author}{\bibfnamefont{I.}~\bibnamefont{Drozdov}},
  \bibinfo{author}{\bibfnamefont{Y.~S.} \bibnamefont{Hor}},
  \bibinfo{author}{\bibfnamefont{R.~J.} \bibnamefont{Cava}}, \bibnamefont{and}
  \bibinfo{author}{\bibfnamefont{A.}~\bibnamefont{Yazdani}},
  \bibinfo{journal}{Nature Physics} \textbf{\bibinfo{volume}{7}},
  \bibinfo{pages}{939} (\bibinfo{year}{2011}).

\bibitem[{\citenamefont{Fu and Kane}(2008)}]{fu_TI_Majorana_2008}
\bibinfo{author}{\bibfnamefont{L.}~\bibnamefont{Fu}} \bibnamefont{and}
  \bibinfo{author}{\bibfnamefont{C.~L.} \bibnamefont{Kane}},
  \bibinfo{journal}{Physical Review Letters} \textbf{\bibinfo{volume}{100}},
  \bibinfo{pages}{096407} (\bibinfo{year}{2008}).

\bibitem[{\citenamefont{Santos et~al.}(2010)\citenamefont{Santos, Neupert,
  Chamon, and Mudry}}]{santos_TI_Majorana_2010}
\bibinfo{author}{\bibfnamefont{L.}~\bibnamefont{Santos}},
  \bibinfo{author}{\bibfnamefont{T.}~\bibnamefont{Neupert}},
  \bibinfo{author}{\bibfnamefont{C.}~\bibnamefont{Chamon}}, \bibnamefont{and}
  \bibinfo{author}{\bibfnamefont{C.}~\bibnamefont{Mudry}},
  \bibinfo{journal}{Physical Review B} \textbf{\bibinfo{volume}{81}},
  \bibinfo{pages}{184502} (\bibinfo{year}{2010}).

\bibitem[{\citenamefont{Qi et~al.}(2010)\citenamefont{Qi, Hughes, and
  Zhang}}]{qi_TI_Majorana_2010}
\bibinfo{author}{\bibfnamefont{X.-L.} \bibnamefont{Qi}},
  \bibinfo{author}{\bibfnamefont{T.~L.} \bibnamefont{Hughes}},
  \bibnamefont{and} \bibinfo{author}{\bibfnamefont{S.-C.} \bibnamefont{Zhang}},
  \bibinfo{journal}{Physical Review B} \textbf{\bibinfo{volume}{82}},
  \bibinfo{pages}{184516} (\bibinfo{year}{2010}).

\bibitem[{\citenamefont{Moore}(2010)}]{moore_birth_2010}
\bibinfo{author}{\bibfnamefont{J.~E.} \bibnamefont{Moore}},
  \bibinfo{journal}{Nature} \textbf{\bibinfo{volume}{464}},
  \bibinfo{pages}{194} (\bibinfo{year}{2010}).

\bibitem[{\citenamefont{Dzero et~al.}(2009)\citenamefont{Dzero, Sun, Galitski,
  and Coleman}}]{Dzero_PRL_2009}
\bibinfo{author}{\bibfnamefont{M.}~\bibnamefont{Dzero}},
  \bibinfo{author}{\bibfnamefont{K.}~\bibnamefont{Sun}},
  \bibinfo{author}{\bibfnamefont{V.}~\bibnamefont{Galitski}}, \bibnamefont{and}
  \bibinfo{author}{\bibfnamefont{P.}~\bibnamefont{Coleman}},
  \bibinfo{journal}{Physical Review Letters} \textbf{\bibinfo{volume}{10}},
  \bibinfo{pages}{106408} (\bibinfo{year}{2009}).

\bibitem[{\citenamefont{Dzero et~al.}(2012)\citenamefont{Dzero, Sun, Coleman,
  and Galitski}}]{Dzero_prb_2012}
\bibinfo{author}{\bibfnamefont{M.}~\bibnamefont{Dzero}},
  \bibinfo{author}{\bibfnamefont{K.}~\bibnamefont{Sun}},
  \bibinfo{author}{\bibfnamefont{P.}~\bibnamefont{Coleman}}, \bibnamefont{and}
  \bibinfo{author}{\bibfnamefont{V.}~\bibnamefont{Galitski}},
  \bibinfo{journal}{Physical Review B} \textbf{\bibinfo{volume}{85}},
  \bibinfo{pages}{045130} (\bibinfo{year}{2012}).

\bibitem[{\citenamefont{Alexandrov
  et~al.}(2013{\natexlab{a}})\citenamefont{Alexandrov, Dzero, and
  Coleman}}]{Dzero_prl_2013}
\bibinfo{author}{\bibfnamefont{V.}~\bibnamefont{Alexandrov}},
  \bibinfo{author}{\bibfnamefont{M.}~\bibnamefont{Dzero}}, \bibnamefont{and}
  \bibinfo{author}{\bibfnamefont{P.}~\bibnamefont{Coleman}},
  \bibinfo{journal}{Physical Review Letters} \textbf{\bibinfo{volume}{111}},
  \bibinfo{pages}{226403} (\bibinfo{year}{2013}{\natexlab{a}}).

\bibitem[{\citenamefont{Ye et~al.}(2013)\citenamefont{Ye, Allen, and
  Sun}}]{sunkai_2013}
\bibinfo{author}{\bibfnamefont{M.}~\bibnamefont{Ye}},
  \bibinfo{author}{\bibfnamefont{J.~W.} \bibnamefont{Allen}}, \bibnamefont{and}
  \bibinfo{author}{\bibfnamefont{K.}~\bibnamefont{Sun}},
  \bibinfo{journal}{arXiv.org} p. \bibinfo{pages}{7191} (\bibinfo{year}{2013}),
  \eprint{1307.7191}.

\bibitem[{\citenamefont{Dzero and Galitski}(2013)}]{Dzero_JETP_2013}
\bibinfo{author}{\bibfnamefont{M.}~\bibnamefont{Dzero}} \bibnamefont{and}
  \bibinfo{author}{\bibfnamefont{V.}~\bibnamefont{Galitski}},
  \bibinfo{journal}{Journal of Experimental and Theoretical Physics}
  \textbf{\bibinfo{volume}{117}}, \bibinfo{pages}{499} (\bibinfo{year}{2013}).

\bibitem[{\citenamefont{Lu et~al.}(2013)\citenamefont{Lu, Zhao, Weng, Fang, and
  Dai}}]{lu_smb6_2013}
\bibinfo{author}{\bibfnamefont{F.}~\bibnamefont{Lu}},
  \bibinfo{author}{\bibfnamefont{J.}~\bibnamefont{Zhao}},
  \bibinfo{author}{\bibfnamefont{H.}~\bibnamefont{Weng}},
  \bibinfo{author}{\bibfnamefont{Z.}~\bibnamefont{Fang}}, \bibnamefont{and}
  \bibinfo{author}{\bibfnamefont{X.}~\bibnamefont{Dai}},
  \bibinfo{journal}{Physical Review Letters} \textbf{\bibinfo{volume}{110}},
  \bibinfo{pages}{096401} (\bibinfo{year}{2013}).

\bibitem[{\citenamefont{Weng et~al.}(2014)\citenamefont{Weng, Zhao, Wang, Fang,
  and Dai}}]{Weng_ybb6_2013}
\bibinfo{author}{\bibfnamefont{H.}~\bibnamefont{Weng}},
  \bibinfo{author}{\bibfnamefont{J.}~\bibnamefont{Zhao}},
  \bibinfo{author}{\bibfnamefont{Z.}~\bibnamefont{Wang}},
  \bibinfo{author}{\bibfnamefont{Z.}~\bibnamefont{Fang}}, \bibnamefont{and}
  \bibinfo{author}{\bibfnamefont{X.}~\bibnamefont{Dai}},
  \bibinfo{journal}{Physical Review Letters} \textbf{\bibinfo{volume}{112}},
  \bibinfo{pages}{016403} (\bibinfo{year}{2014}).

\bibitem[{\citenamefont{Deng et~al.}(2013)\citenamefont{Deng, Haule, and
  Kotliar}}]{Deng_TKI_2013}
\bibinfo{author}{\bibfnamefont{X.}~\bibnamefont{Deng}},
  \bibinfo{author}{\bibfnamefont{K.}~\bibnamefont{Haule}}, \bibnamefont{and}
  \bibinfo{author}{\bibfnamefont{G.}~\bibnamefont{Kotliar}},
  \bibinfo{journal}{Physical Review Letters} \textbf{\bibinfo{volume}{111}},
  \bibinfo{pages}{176404} (\bibinfo{year}{2013}).

\bibitem[{\citenamefont{Alexandrov
  et~al.}(2013{\natexlab{b}})\citenamefont{Alexandrov, Dzero, and
  Coleman}}]{Coleman_2013PRL}
\bibinfo{author}{\bibfnamefont{V.}~\bibnamefont{Alexandrov}},
  \bibinfo{author}{\bibfnamefont{M.}~\bibnamefont{Dzero}}, \bibnamefont{and}
  \bibinfo{author}{\bibfnamefont{P.}~\bibnamefont{Coleman}},
  \bibinfo{journal}{Physical Review Letters} \textbf{\bibinfo{volume}{111}},
  \bibinfo{pages}{226403} (\bibinfo{year}{2013}{\natexlab{b}}).

\bibitem[{\citenamefont{Legner et~al.}(2014)\citenamefont{Legner, Ruegg, and
  Sigrist}}]{MSigrist_2014_TKI}
\bibinfo{author}{\bibfnamefont{M.}~\bibnamefont{Legner}},
  \bibinfo{author}{\bibfnamefont{A.}~\bibnamefont{Ruegg}}, \bibnamefont{and}
  \bibinfo{author}{\bibfnamefont{M.}~\bibnamefont{Sigrist}},
  \bibinfo{journal}{Physical Review B} \textbf{\bibinfo{volume}{89}},
  \bibinfo{pages}{085110} (\bibinfo{year}{2014}).

\bibitem[{\citenamefont{Werner and Assaad}(2013)}]{Werner_TKI_2013}
\bibinfo{author}{\bibfnamefont{J.}~\bibnamefont{Werner}} \bibnamefont{and}
  \bibinfo{author}{\bibfnamefont{F.~F.} \bibnamefont{Assaad}},
  \bibinfo{journal}{Physical Review B} \textbf{\bibinfo{volume}{88}},
  \bibinfo{pages}{035113} (\bibinfo{year}{2013}).

\bibitem[{\citenamefont{Frantzeskakis
  et~al.}(2013{\natexlab{a}})\citenamefont{Frantzeskakis, de~Jong,
  Zwartsenberg, Huang, Pan, Zhang, Zhang, Zhang, Bao, Tegus
  et~al.}}]{smb6_arpes_Frantzeskakis_2013}
\bibinfo{author}{\bibfnamefont{E.}~\bibnamefont{Frantzeskakis}},
  \bibinfo{author}{\bibfnamefont{N.}~\bibnamefont{de~Jong}},
  \bibinfo{author}{\bibfnamefont{B.}~\bibnamefont{Zwartsenberg}},
  \bibinfo{author}{\bibfnamefont{Y.}~\bibnamefont{Huang}},
  \bibinfo{author}{\bibfnamefont{Y.}~\bibnamefont{Pan}},
  \bibinfo{author}{\bibfnamefont{X.}~\bibnamefont{Zhang}},
  \bibinfo{author}{\bibfnamefont{J.}~\bibnamefont{Zhang}},
  \bibinfo{author}{\bibfnamefont{F.}~\bibnamefont{Zhang}},
  \bibinfo{author}{\bibfnamefont{L.}~\bibnamefont{Bao}},
  \bibinfo{author}{\bibfnamefont{O.}~\bibnamefont{Tegus}},
  \bibnamefont{et~al.}, \bibinfo{journal}{Physical Review X}
  \textbf{\bibinfo{volume}{3}}, \bibinfo{pages}{041024}
  (\bibinfo{year}{2013}{\natexlab{a}}).

\bibitem[{\citenamefont{Wolgast et~al.}(2013)\citenamefont{Wolgast, Kurdak,
  Sun, Allen, Kim, and Fisk}}]{smb6_ss_conduct_2013}
\bibinfo{author}{\bibfnamefont{S.}~\bibnamefont{Wolgast}},
  \bibinfo{author}{\bibfnamefont{{\c C}.}~\bibnamefont{Kurdak}},
  \bibinfo{author}{\bibfnamefont{K.}~\bibnamefont{Sun}},
  \bibinfo{author}{\bibfnamefont{J.~W.} \bibnamefont{Allen}},
  \bibinfo{author}{\bibfnamefont{D.-J.} \bibnamefont{Kim}}, \bibnamefont{and}
  \bibinfo{author}{\bibfnamefont{Z.}~\bibnamefont{Fisk}},
  \bibinfo{journal}{Physical Review B} \textbf{\bibinfo{volume}{88}},
  \bibinfo{pages}{180405} (\bibinfo{year}{2013}).

\bibitem[{\citenamefont{Xu et~al.}(2013)\citenamefont{Xu, Shi, Biswas, Matt,
  Dhaka, Huang, Plumb, Radovic, Dil, Pomjakushina
  et~al.}}]{smb6_arpes_xunan_2013}
\bibinfo{author}{\bibfnamefont{N.}~\bibnamefont{Xu}},
  \bibinfo{author}{\bibfnamefont{X.}~\bibnamefont{Shi}},
  \bibinfo{author}{\bibfnamefont{P.~K.} \bibnamefont{Biswas}},
  \bibinfo{author}{\bibfnamefont{C.~E.} \bibnamefont{Matt}},
  \bibinfo{author}{\bibfnamefont{R.~S.} \bibnamefont{Dhaka}},
  \bibinfo{author}{\bibfnamefont{Y.}~\bibnamefont{Huang}},
  \bibinfo{author}{\bibfnamefont{N.~C.} \bibnamefont{Plumb}},
  \bibinfo{author}{\bibfnamefont{M.}~\bibnamefont{Radovic}},
  \bibinfo{author}{\bibfnamefont{J.~H.} \bibnamefont{Dil}},
  \bibinfo{author}{\bibfnamefont{E.}~\bibnamefont{Pomjakushina}},
  \bibnamefont{et~al.}, \bibinfo{journal}{Physical Review B}
  \textbf{\bibinfo{volume}{88}}, \bibinfo{pages}{121102}
  (\bibinfo{year}{2013}).

\bibitem[{\citenamefont{Li et~al.}(2013)\citenamefont{Li, Xiang, Yu, Asaba,
  Lawson, Cai, Tinsman, Berkley, Wolgast, Eo et~al.}}]{smb6_qoscillation_2013}
\bibinfo{author}{\bibfnamefont{G.}~\bibnamefont{Li}},
  \bibinfo{author}{\bibfnamefont{Z.}~\bibnamefont{Xiang}},
  \bibinfo{author}{\bibfnamefont{F.}~\bibnamefont{Yu}},
  \bibinfo{author}{\bibfnamefont{T.}~\bibnamefont{Asaba}},
  \bibinfo{author}{\bibfnamefont{B.}~\bibnamefont{Lawson}},
  \bibinfo{author}{\bibfnamefont{P.}~\bibnamefont{Cai}},
  \bibinfo{author}{\bibfnamefont{C.}~\bibnamefont{Tinsman}},
  \bibinfo{author}{\bibfnamefont{A.}~\bibnamefont{Berkley}},
  \bibinfo{author}{\bibfnamefont{S.}~\bibnamefont{Wolgast}},
  \bibinfo{author}{\bibfnamefont{Y.~S.} \bibnamefont{Eo}},
  \bibnamefont{et~al.}, \bibinfo{journal}{arXiv.org} p. \bibinfo{pages}{5221}
  (\bibinfo{year}{2013}), \eprint{1306.5221}.

\bibitem[{\citenamefont{Neupane et~al.}(2013)\citenamefont{Neupane, Alidoust,
  Xu, Kondo, Ishida, Kim, Liu, Belopolski, Jo, Chang
  et~al.}}]{smb6_arpes_Neupane_2013}
\bibinfo{author}{\bibfnamefont{M.}~\bibnamefont{Neupane}},
  \bibinfo{author}{\bibfnamefont{N.}~\bibnamefont{Alidoust}},
  \bibinfo{author}{\bibfnamefont{S.~Y.} \bibnamefont{Xu}},
  \bibinfo{author}{\bibfnamefont{T.}~\bibnamefont{Kondo}},
  \bibinfo{author}{\bibfnamefont{Y.}~\bibnamefont{Ishida}},
  \bibinfo{author}{\bibfnamefont{D.~J.} \bibnamefont{Kim}},
  \bibinfo{author}{\bibfnamefont{C.}~\bibnamefont{Liu}},
  \bibinfo{author}{\bibfnamefont{I.}~\bibnamefont{Belopolski}},
  \bibinfo{author}{\bibfnamefont{Y.~J.} \bibnamefont{Jo}},
  \bibinfo{author}{\bibfnamefont{T.~R.} \bibnamefont{Chang}},
  \bibnamefont{et~al.}, \bibinfo{journal}{Nature Communications}
  \textbf{\bibinfo{volume}{4}} (\bibinfo{year}{2013}).

\bibitem[{\citenamefont{Kim et~al.}(2013)\citenamefont{Kim, Thomas, Grant,
  Botimer, Fisk, and Xia}}]{smb6_Transport_2013}
\bibinfo{author}{\bibfnamefont{D.~J.} \bibnamefont{Kim}},
  \bibinfo{author}{\bibfnamefont{S.}~\bibnamefont{Thomas}},
  \bibinfo{author}{\bibfnamefont{T.}~\bibnamefont{Grant}},
  \bibinfo{author}{\bibfnamefont{J.}~\bibnamefont{Botimer}},
  \bibinfo{author}{\bibfnamefont{Z.}~\bibnamefont{Fisk}}, \bibnamefont{and}
  \bibinfo{author}{\bibfnamefont{J.}~\bibnamefont{Xia}},
  \bibinfo{journal}{Nature Scientific Reports} \textbf{\bibinfo{volume}{3}},
  \bibinfo{pages}{3150} (\bibinfo{year}{2013}).

\bibitem[{\citenamefont{Jiang et~al.}(2013)\citenamefont{Jiang, Li, Zhang, Sun,
  Chen, Ye, Xu, Ge, Tan, Niu et~al.}}]{smb6_arpes_fengdl_2013}
\bibinfo{author}{\bibfnamefont{J.}~\bibnamefont{Jiang}},
  \bibinfo{author}{\bibfnamefont{S.}~\bibnamefont{Li}},
  \bibinfo{author}{\bibfnamefont{T.}~\bibnamefont{Zhang}},
  \bibinfo{author}{\bibfnamefont{Z.}~\bibnamefont{Sun}},
  \bibinfo{author}{\bibfnamefont{F.}~\bibnamefont{Chen}},
  \bibinfo{author}{\bibfnamefont{Z.~R.} \bibnamefont{Ye}},
  \bibinfo{author}{\bibfnamefont{M.}~\bibnamefont{Xu}},
  \bibinfo{author}{\bibfnamefont{Q.~Q.} \bibnamefont{Ge}},
  \bibinfo{author}{\bibfnamefont{S.~Y.} \bibnamefont{Tan}},
  \bibinfo{author}{\bibfnamefont{X.~H.} \bibnamefont{Niu}},
  \bibnamefont{et~al.}, \bibinfo{journal}{Nature Communications}
  \textbf{\bibinfo{volume}{4}}, \bibinfo{pages}{3010} (\bibinfo{year}{2013}).

\bibitem[{\citenamefont{Thomas et~al.}(2013)\citenamefont{Thomas, Kim, Chung,
  Grant, Fisk, and Xia}}]{smb6_WAL_2013}
\bibinfo{author}{\bibfnamefont{S.}~\bibnamefont{Thomas}},
  \bibinfo{author}{\bibfnamefont{D.~J.} \bibnamefont{Kim}},
  \bibinfo{author}{\bibfnamefont{S.~B.} \bibnamefont{Chung}},
  \bibinfo{author}{\bibfnamefont{T.}~\bibnamefont{Grant}},
  \bibinfo{author}{\bibfnamefont{Z.}~\bibnamefont{Fisk}}, \bibnamefont{and}
  \bibinfo{author}{\bibfnamefont{J.}~\bibnamefont{Xia}},
  \bibinfo{journal}{arXiv.org} p. \bibinfo{pages}{4133} (\bibinfo{year}{2013}),
  \eprint{1307.4133}.

\bibitem[{\citenamefont{Yee et~al.}(2013)\citenamefont{Yee, He,
  Soumyanarayanan, Kim, Fisk, and Hoffman}}]{smb6_Yee_2013}
\bibinfo{author}{\bibfnamefont{M.~M.} \bibnamefont{Yee}},
  \bibinfo{author}{\bibfnamefont{Y.}~\bibnamefont{He}},
  \bibinfo{author}{\bibfnamefont{A.}~\bibnamefont{Soumyanarayanan}},
  \bibinfo{author}{\bibfnamefont{D.-J.} \bibnamefont{Kim}},
  \bibinfo{author}{\bibfnamefont{Z.}~\bibnamefont{Fisk}}, \bibnamefont{and}
  \bibinfo{author}{\bibfnamefont{J.~E.} \bibnamefont{Hoffman}},
  \bibinfo{journal}{arXiv.org} p. \bibinfo{pages}{1085} (\bibinfo{year}{2013}),
  \eprint{1308.1085}.

\bibitem[{\citenamefont{Denlinger
  et~al.}(2013{\natexlab{a}})\citenamefont{Denlinger, Allen, Kang, Sun, Kim,
  Shim, Min, Kim, and Fisk}}]{exp_TKI_2013}
\bibinfo{author}{\bibfnamefont{J.~D.} \bibnamefont{Denlinger}},
  \bibinfo{author}{\bibfnamefont{J.~W.} \bibnamefont{Allen}},
  \bibinfo{author}{\bibfnamefont{J.~S.} \bibnamefont{Kang}},
  \bibinfo{author}{\bibfnamefont{K.}~\bibnamefont{Sun}},
  \bibinfo{author}{\bibfnamefont{J.~W.} \bibnamefont{Kim}},
  \bibinfo{author}{\bibfnamefont{J.~H.} \bibnamefont{Shim}},
  \bibinfo{author}{\bibfnamefont{B.~I.} \bibnamefont{Min}},
  \bibinfo{author}{\bibfnamefont{D.-J.} \bibnamefont{Kim}}, \bibnamefont{and}
  \bibinfo{author}{\bibfnamefont{Z.}~\bibnamefont{Fisk}},
  \bibinfo{journal}{arXiv.org} p. \bibinfo{pages}{6637}
  (\bibinfo{year}{2013}{\natexlab{a}}), \eprint{1312.6637}.

\bibitem[{\citenamefont{Frantzeskakis
  et~al.}(2013{\natexlab{b}})\citenamefont{Frantzeskakis, de~Jong,
  Zwartsenberg, Huang, Pan, Zhang, Zhang, Zhang, Bao, Tegus
  et~al.}}]{exp_ss_PRX_2013}
\bibinfo{author}{\bibfnamefont{E.}~\bibnamefont{Frantzeskakis}},
  \bibinfo{author}{\bibfnamefont{N.}~\bibnamefont{de~Jong}},
  \bibinfo{author}{\bibfnamefont{B.}~\bibnamefont{Zwartsenberg}},
  \bibinfo{author}{\bibfnamefont{Y.~K.} \bibnamefont{Huang}},
  \bibinfo{author}{\bibfnamefont{Y.}~\bibnamefont{Pan}},
  \bibinfo{author}{\bibfnamefont{X.}~\bibnamefont{Zhang}},
  \bibinfo{author}{\bibfnamefont{J.~X.} \bibnamefont{Zhang}},
  \bibinfo{author}{\bibfnamefont{F.~X.} \bibnamefont{Zhang}},
  \bibinfo{author}{\bibfnamefont{L.~H.} \bibnamefont{Bao}},
  \bibinfo{author}{\bibfnamefont{O.}~\bibnamefont{Tegus}},
  \bibnamefont{et~al.}, \bibinfo{journal}{Physical Review X}
  \textbf{\bibinfo{volume}{3}}, \bibinfo{pages}{4260431}
  (\bibinfo{year}{2013}{\natexlab{b}}).

\bibitem[{\citenamefont{Ruan et~al.}(2014)\citenamefont{Ruan, Ye, Guo, Chen,
  Chen, Zhang, and Wang}}]{SmB6_stm_2014PRL}
\bibinfo{author}{\bibfnamefont{W.}~\bibnamefont{Ruan}},
  \bibinfo{author}{\bibfnamefont{C.}~\bibnamefont{Ye}},
  \bibinfo{author}{\bibfnamefont{M.}~\bibnamefont{Guo}},
  \bibinfo{author}{\bibfnamefont{F.}~\bibnamefont{Chen}},
  \bibinfo{author}{\bibfnamefont{X.}~\bibnamefont{Chen}},
  \bibinfo{author}{\bibfnamefont{G.-M.} \bibnamefont{Zhang}}, \bibnamefont{and}
  \bibinfo{author}{\bibfnamefont{Y.}~\bibnamefont{Wang}},
  \bibinfo{journal}{Physical Review Letters} \textbf{\bibinfo{volume}{112}},
  \bibinfo{pages}{136401} (\bibinfo{year}{2014}).

\bibitem[{\citenamefont{Denlinger
  et~al.}(2013{\natexlab{b}})\citenamefont{Denlinger, Allen, Kang, Sun, Min,
  Kim, and Fisk}}]{ARPES_smb6_review_2013}
\bibinfo{author}{\bibfnamefont{J.~D.} \bibnamefont{Denlinger}},
  \bibinfo{author}{\bibfnamefont{J.~W.} \bibnamefont{Allen}},
  \bibinfo{author}{\bibfnamefont{J.-S.} \bibnamefont{Kang}},
  \bibinfo{author}{\bibfnamefont{K.}~\bibnamefont{Sun}},
  \bibinfo{author}{\bibfnamefont{B.-I.} \bibnamefont{Min}},
  \bibinfo{author}{\bibfnamefont{D.-J.} \bibnamefont{Kim}}, \bibnamefont{and}
  \bibinfo{author}{\bibfnamefont{Z.}~\bibnamefont{Fisk}},
  \bibinfo{journal}{arXiv.org} p. \bibinfo{pages}{6636}
  (\bibinfo{year}{2013}{\natexlab{b}}), \eprint{1312.6636}.

\bibitem[{\citenamefont{Beaurepaire et~al.}(1990)\citenamefont{Beaurepaire,
  Kappler, and Krill}}]{beaurepaire_MVS_1990}
\bibinfo{author}{\bibfnamefont{E.}~\bibnamefont{Beaurepaire}},
  \bibinfo{author}{\bibfnamefont{J.~P.} \bibnamefont{Kappler}},
  \bibnamefont{and} \bibinfo{author}{\bibfnamefont{G.}~\bibnamefont{Krill}},
  \bibinfo{journal}{Physical Review B} \textbf{\bibinfo{volume}{41}},
  \bibinfo{pages}{6768} (\bibinfo{year}{1990}).

\bibitem[{\citenamefont{Cohen et~al.}(1970)\citenamefont{Cohen, Eibschutz, and
  West}}]{cohen_SmB6_Kondo_1970}
\bibinfo{author}{\bibfnamefont{R.~L.} \bibnamefont{Cohen}},
  \bibinfo{author}{\bibfnamefont{M.}~\bibnamefont{Eibschutz}},
  \bibnamefont{and} \bibinfo{author}{\bibfnamefont{K.~W.} \bibnamefont{West}},
  \bibinfo{journal}{Physical Review Letters} \textbf{\bibinfo{volume}{24}},
  \bibinfo{pages}{383} (\bibinfo{year}{1970}).

\bibitem[{\citenamefont{Eibschutz et~al.}(1972)\citenamefont{Eibschutz, Cohen,
  Buehler, and Wernick}}]{eibschutz_SmB6_1972}
\bibinfo{author}{\bibfnamefont{M.}~\bibnamefont{Eibschutz}},
  \bibinfo{author}{\bibfnamefont{R.~L.} \bibnamefont{Cohen}},
  \bibinfo{author}{\bibfnamefont{E.}~\bibnamefont{Buehler}}, \bibnamefont{and}
  \bibinfo{author}{\bibfnamefont{J.~H.} \bibnamefont{Wernick}},
  \bibinfo{journal}{Physical Review B} \textbf{\bibinfo{volume}{6}},
  \bibinfo{pages}{18} (\bibinfo{year}{1972}).

\bibitem[{\citenamefont{Chazalviel et~al.}(1976)\citenamefont{Chazalviel,
  Campagna, Wertheim, and Schmidt}}]{chazalviel_MVs_1976}
\bibinfo{author}{\bibfnamefont{J.~N.} \bibnamefont{Chazalviel}},
  \bibinfo{author}{\bibfnamefont{M.}~\bibnamefont{Campagna}},
  \bibinfo{author}{\bibfnamefont{G.~K.} \bibnamefont{Wertheim}},
  \bibnamefont{and} \bibinfo{author}{\bibfnamefont{P.~H.}
  \bibnamefont{Schmidt}}, \bibinfo{journal}{Physical Review B}
  \textbf{\bibinfo{volume}{14}}, \bibinfo{pages}{4586} (\bibinfo{year}{1976}).

\bibitem[{\citenamefont{Kane and Mele}(2005)}]{kane_z2_2005}
\bibinfo{author}{\bibfnamefont{C.~L.} \bibnamefont{Kane}} \bibnamefont{and}
  \bibinfo{author}{\bibfnamefont{E.~J.} \bibnamefont{Mele}},
  \bibinfo{journal}{Physical Review Letters} \textbf{\bibinfo{volume}{95}},
  \bibinfo{pages}{146802} (\bibinfo{year}{2005}).

\bibitem[{\citenamefont{Fu and Kane}(2006)}]{Fu_kane_z2_2006}
\bibinfo{author}{\bibfnamefont{L.}~\bibnamefont{Fu}} \bibnamefont{and}
  \bibinfo{author}{\bibfnamefont{C.}~\bibnamefont{Kane}},
  \bibinfo{journal}{Physical Review B} \textbf{\bibinfo{volume}{74}},
  \bibinfo{pages}{195312} (\bibinfo{year}{2006}).

\bibitem[{\citenamefont{Deng et~al.}(2009)\citenamefont{Deng, Wang, Dai, and
  Fang}}]{DengXY_GW}
\bibinfo{author}{\bibfnamefont{X.}~\bibnamefont{Deng}},
  \bibinfo{author}{\bibfnamefont{L.}~\bibnamefont{Wang}},
  \bibinfo{author}{\bibfnamefont{X.}~\bibnamefont{Dai}}, \bibnamefont{and}
  \bibinfo{author}{\bibfnamefont{Z.}~\bibnamefont{Fang}},
  \bibinfo{journal}{Physical Review B} \textbf{\bibinfo{volume}{79}},
  \bibinfo{pages}{075114} (\bibinfo{year}{2009}).

\bibitem[{\citenamefont{Kotliar et~al.}(2006)\citenamefont{Kotliar, Savrasov,
  Haule, Oudovenko, Parcollet, and Marianetti}}]{Kotliar:2006}
\bibinfo{author}{\bibfnamefont{G.}~\bibnamefont{Kotliar}},
  \bibinfo{author}{\bibfnamefont{S.}~\bibnamefont{Savrasov}},
  \bibinfo{author}{\bibfnamefont{K.}~\bibnamefont{Haule}},
  \bibinfo{author}{\bibfnamefont{V.}~\bibnamefont{Oudovenko}},
  \bibinfo{author}{\bibfnamefont{O.}~\bibnamefont{Parcollet}},
  \bibnamefont{and}
  \bibinfo{author}{\bibfnamefont{C.}~\bibnamefont{Marianetti}},
  \bibinfo{journal}{Reviews of Modern Physics} \textbf{\bibinfo{volume}{78}},
  \bibinfo{pages}{865} (\bibinfo{year}{2006}).

\bibitem[{\citenamefont{Kuneš et~al.}(2010)\citenamefont{Kuneš, Arita,
  Wissgott, Toschi, Ikeda, and Held}}]{wannier_2010}
\bibinfo{author}{\bibfnamefont{J.}~\bibnamefont{Kuneš}},
  \bibinfo{author}{\bibfnamefont{R.}~\bibnamefont{Arita}},
  \bibinfo{author}{\bibfnamefont{P.}~\bibnamefont{Wissgott}},
  \bibinfo{author}{\bibfnamefont{A.}~\bibnamefont{Toschi}},
  \bibinfo{author}{\bibfnamefont{H.}~\bibnamefont{Ikeda}}, \bibnamefont{and}
  \bibinfo{author}{\bibfnamefont{K.}~\bibnamefont{Held}},
  \bibinfo{journal}{Computer Physics Communications}
  \textbf{\bibinfo{volume}{181}}, \bibinfo{pages}{1888} (\bibinfo{year}{2010}).

\bibitem[{\citenamefont{Solovyev et~al.}(2007)\citenamefont{Solovyev,
  Pchelkina, and Anisimov}}]{wannier_2007}
\bibinfo{author}{\bibfnamefont{I.~V.} \bibnamefont{Solovyev}},
  \bibinfo{author}{\bibfnamefont{Z.~V.} \bibnamefont{Pchelkina}},
  \bibnamefont{and} \bibinfo{author}{\bibfnamefont{V.~I.}
  \bibnamefont{Anisimov}}, \bibinfo{journal}{Physical Review B}
  \textbf{\bibinfo{volume}{75}}, \bibinfo{pages}{045110}
  (\bibinfo{year}{2007}).

\bibitem[{\citenamefont{Mostofi et~al.}(2008)\citenamefont{Mostofi, Yates, Lee,
  Souza, Vanderbilt, and Marzari}}]{wannier_2008}
\bibinfo{author}{\bibfnamefont{A.~A.} \bibnamefont{Mostofi}},
  \bibinfo{author}{\bibfnamefont{J.~R.} \bibnamefont{Yates}},
  \bibinfo{author}{\bibfnamefont{Y.-S.} \bibnamefont{Lee}},
  \bibinfo{author}{\bibfnamefont{I.}~\bibnamefont{Souza}},
  \bibinfo{author}{\bibfnamefont{D.}~\bibnamefont{Vanderbilt}},
  \bibnamefont{and} \bibinfo{author}{\bibfnamefont{N.}~\bibnamefont{Marzari}},
  \bibinfo{journal}{Computer Physics Communications}
  \textbf{\bibinfo{volume}{178}}, \bibinfo{pages}{685} (\bibinfo{year}{2008}).

\end{thebibliography}

\end{document}